# SiC/YAG composite coatings by a novel liquid fuelled high velocity oxy-fuel suspension thermal spray

F. Venturi*, A. Rincón Romero, T. Hussain

**Affiliations:**

Faculty of Engineering, University of Nottingham, University Park, Nottingham, NG7 2RD, United Kingdom

*Corresponding author: federico.venturi@nottingham.ac.uk

**Abstract**

Despite recent advances in suspension-based thermal spray techniques, there is still a need to widen the capability of available thermal spray setups to handle suspension to explore new compositions and improve the properties and performance of existing ones. In this work, a novel setup for injecting liquid-based feedstock in a liquid-fuelled high velocity oxy-fuel (HVOLF) thermal spray torch is proposed, involving a new hardware ("Hybrid Nozzle") capable of injecting suspension radially and protecting the thermal spray flame from the environmental oxygen by inert gas shrouding. The new setup was used to deposit SiC/YAG composite coatings with 85 wt. % SiC content, and their wear performance was evaluated. An oxygen deficient environment was provided to the HVOLF flame, ensuring no or minimal (<2 wt. %) $SiO_2$ was formed in the coating during the spray, as confirmed by XRD. A good wear performance was observed at low load below 20 N, with a specific wear rate < 5 x $10^{-5}$ $mm^3$/Nm, with the onset of wear mechanisms characteristic of ceramic wear in the brittle regime between 20 and 30 N. The future applicability of this novel setup to a range of oxidation- and heat-sensitive materials and composites offers the opportunity for new coating materials to be explored with suspension and solution precursor-based HVOLF thermal spray.

## 1 - Introduction

Silicon carbide (SiC) is a ceramic widely used for its superior mechanical properties, making it an interesting, exciting material with applications e.g. in machining as a cutting tool, automotive as brake discs, and as structural material for ballistic vests. The use of SiC as a coating is extensive, especially for its tribological properties of low friction and low wear [1], which make it ideal for wear resistant applications. Applications in coating form include safety critical parts such as offshore wind turbine bearings [2]. SiC is generally produced from silica and carbon by the Acheson process [3], and more recent developments have shown to produce finer, higher purity SiC from high purity silica and carbon black at high temperature [4]. Among consolidating methods, starting from raw SiC and producing bulk SiC, it is worth mentioning the Lely process [5] and spark plasma sintering [6].

The situation becomes more challenging when aiming at depositing SiC coatings, due to the tendency of SiC at high temperature to sublimate [7], decompose and subsequently oxidise to form $SiO_2$ [8, 9]. These processes are ultimately detrimental to the coating quality and performance. To prevent SiC particles from sublimation and oxidation when depositing coatings, a sintering aid or binder material is often added, which ultimately yields composite coatings with varying SiC content. The binder, as the name suggests, also has the role of holding together the SiC particles. For instance, the combination of SiC with $Al_2O_3$ and $Y_2O_3$ deposited by plasma spray in the atmosphere has shown to provide SiC coatings with minimal oxidation, however with poorer mechanical properties compared to bulk SiC due to weak bonding, or cohesion, between the constituent particles in the coating [10]. An additional measure to tackle the oxidation issue has been to deposit the coatings in an oxygen-depleted environment to prevent oxidation as a result of the high temperature, where SiC is more reactive. A combination of these two approaches has been done by using plasma spray in controlled atmosphere (argon gas) along with $ZrB_2$ [11]. Another approach used for plasma spray in a vacuum is the deposition of Si coatings on C/C substrates to generate SiC in a subsequent heat treatment [12]. A similar approach was used to spray $Si(OH)_4$ liquid precursor on C/C, forming a SiC interlayer between $SiO_2$ and C/C [13].

Alternative spray approaches are cold spray and flame spray. In cold spray, there is little risk of SiC oxidation, but a soft binder metal like aluminium is needed to obtain the deposition [14]. In flame spray, it was possible to achieve little SiC oxidation, however with low SiC content (< 40 wt. %) together with a NiCrSiBFeC primary phase [15]. Recent advances in Suspension High Velocity Oxy-Fuel (SHVOF) thermal spray have allowed to deposit SiC/YAG coatings with no or minimal oxidation [16] using a hydrogen-fuelled GTV Top Gun SHVOF torch. The suspension route is a very interesting approach, worth exploring due to the simplicity in handling finer particles, solution precursors and combinations of these and dissimilar materials, including for SiC coatings [17]. Also very interesting is the use of liquid-fuelled HVOF (referred to as HVOLF), which provides a higher flame power yielding greater velocity and lower temperature than its hydrogen-based counterparts, proving even more beneficial for oxidation sensitive materials. However, not all HVOF torches are designed to allow suspension injection, and in particular, none of the liquid-fuelled torches currently is.

To avoid the limitations and technical challenges of spraying in a controlled-atmosphere enclosure, such as in vacuum plasma spray, gas shrouds have been proposed [18], with the purpose of avoiding mixing of the oxygen from the environment with the thermal spray flame. Further modelling works [19] have shown that a physical shrouding of the HVOF flame has a greater effect of hindering oxygen entrainment and oxygen present in the coating than the flowing of gas itself. Being oxidation and heat-degradation in general aided by the high temperature of the process, another approach to this issue is radial injection. Radial injection, customarily used in plasma spray, has been used in HVOF thermal spray to handle sensitive materials such as titanium oxide [20] to avoid unwanted anatase to rutile phase transformation, and graphene [21] to hinder oxidation to graphene oxide or complete degradation. In particular, it has been shown how the angle of incidence, position and flow rate of the radial injection, affect the feedstock temperature and velocity at the various flame power values, suggesting a way to tailor the flame-feedstock interaction in order to achieve the desired particle temperature and velocity [22]. Currently available solutions [18-20] therefore focus on only one of these aspects at a time, and are not compatible with higher flame power, liquid-fuelled thermal spray systems. The effect of the combination of physical shrouding, gas shrouding and radial injection on the feedstock has been recently studied by simulation [23]. Radial injection alone is feasible in hydrogen-fuelled HVOF thermal spray torches, which normally reach around 100 kW flame power, but a desirable flame penetration by the feedstock is challenging for liquid-fuelled HVOF torches capable of reaching 250 kW flame power. Therefore, the coupling of a physical shroud and radial injection can allow a suitable flame penetration since the feedstock injection happens within the physical shroud, with no need to overtake the flame boundaries.

The search for an efficient and scalable means of deposition for coating materials subject to heat- and oxygen-related degradation, such as carbides, nitrides, nanomaterials and pure metals, is still ongoing. Therefore, new techniques or combinations of existing approaches are needed. In this work, we present a novel setup for HVOLF that includes the use of a solid shroud, a shrouding inert gas, radial feedstock injection, and the use of suspension feedstock of SiC/YAG composite particles. The aim of this work is to prove the use of the newly developed hardware and analyse one coating deposition as a case study. The setup endows for the first time the MetJet4 HVOLF thermal spray torch (Metallisation, UK) the capability of handling liquid feedstock such as suspension or solution precursor. Coatings with no or minimal (< 2 wt. %) SiC oxidation detectable through X-Ray diffraction were obtained. The mechanical and tribological properties of the coating were analysed, and the potential applicability of the setup is discussed. This work opens the way to potential application to form new composite coatings with heat- oxidation-sensitive materials (e.g. nanocomposites containing graphene and boron nitride) to enhance tribological properties of wear resistant coatings.

## 2 - Methods

### 2.1 - Feedstock materials

The feedstock used was a commercial SiC/YAG powder (Seram Coatings, Norway), agglomerated and sintered, whose cross-sectional Scanning Electron Microscopy (SEM) images are shown in Figure 1. Irregular round-shaped particles can be seen, with partially hollow cores and an intermixing of SiC grains (grey phase) and YAG binder (bright phase). The particle size is in the range 15-45 um and the composition is approximately 70 wt. % SiC and 30 wt. % YAG, as stated by the manufacturer. A water-based suspension was prepared from this powder, yielding a solid load of 20 wt. %. The stability of the suspension was ensured by adding 0.2 wt. % polyacrylic acid sodium salt dispersant (Sigma Aldrich, UK), adjusting the pH to 9, and stirring mechanically and ultrasonically, according to the procedure presented in [16].

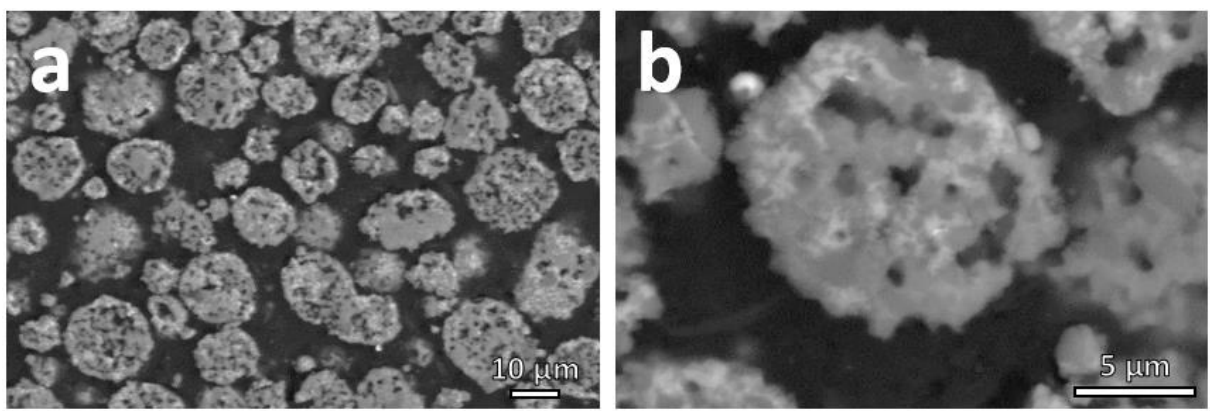

**Figure 1 – SiC/YAG feedstock –** Low (a) and high (b) magnification BSE SEM images of the cross-sectioned SiC/YAG powder.

### 2.2 - Experimental setup

High velocity oxy-fuel thermal spray was carried out with a liquid fuelled system MetJet 4 (Metallisation Ltd, United Kingdom). The kerosene flow was 500 ml/min, and the oxygen flow 900 l/min, yielding a stoichiometry of 91% and a total mass flow of 28 g/s. The choice of this stoichiometry is justified by the need of leaving a small fraction of the fuel unburnt in order leave no free oxygen available for oxidising the feedstock material. The substrates were 60 x 25 x 2 mm coupons of AISI304 stainless steel (nominal composition: 18% Cr, 8% Ni, 2% Mn, 0.08% C, 0.045% P, 0.03% S, 0.75% Si, 0.1% N, all in wt. %, and Fe balance). The substrates were grit blasted with SiC grit at 6 bar and ultrasonicated in deionised water and industrial methylated spirit mixture. The substrates were positioned at a stand-off distance of 203 mm (6 inches). The coating was sprayed in a raster path with 4 mm overlap, a traverse speed of 1 m/s and for a total of 20 passes.

The modified setup of this work includes an additional attachment, the "Hybrid Nozzle", placed on the thermal spray torch as schematised in Figure 2. The Hybrid Nozzle provides the radial injection of suspension feedstock as well as the injection of an inert gas ($N_2$ in this case) to prevent oxygen entrainment from the atmosphere into the thermal spray jet. The suspension feedstock was injected at 50 ml/min using a feedback loop-controlled pressurised vessel. Compared to a powder feedstock spray, part of the flame power is lost for water vaporisation, but this is just a small fraction at this feedstock flowrate (1.88 kW). In this setup, no feedstock is injected axially in the original thermal spray torch; however, $N_2$ carrier gas was still fluxed at 9 ml/min to prevent the hot jet from coming back in the feeder tubes. The detailed design of the Hybrid Nozzle is confidential and cannot be shared at this stage, but the schematic in Figure 2 shows the working principle and the active elements without the dimensions.

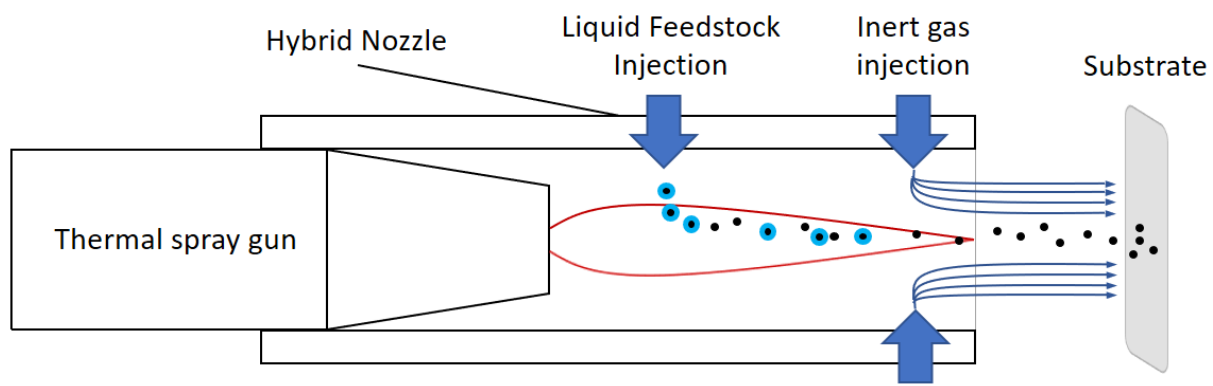

**Figure 2 – Experimental setup –** The Hybrid Nozzle is attached to a HVOLF thermal spray torch, providing radial suspension injection and inert gas shrouding of the thermal spray jet and feedstock.

**2.3 - Sample characterisation and wear testing**

The porosity of the coating was measured with ImageJ (NIH, USA) using the thresholding technique on twenty 100 µm² contiguous areas on the polished cross-section, and average values with associated standard error were reported. The microhardness of the coating was measured using a Microhardness tester (Buehler, UK) with 10 indents at the mid-thickness of the cross section, with a 25 gf load chosen as appropriate for the coating thickness. The associated error is the standard error. The SEM images were acquired using a 6490LV SEM (JEOL, Japan) and an XL30 SEM (FEI, The Netherlands) at 20 kV accelerating voltage. Using the JEOL 6490LV, Wavelength Dispersive X-ray Spectroscopy (WDS) data has been acquired with an Oxford Instruments INCA Wave 700 WDS System for high resolution elemental quantification, calibrated with Si, O and C references prior to each WDS scan. The X-ray diffractograms were acquired using a D8 Advance Da Vinci (Bruker, Germany) with Cu cathode ($K_\alpha \lambda$ = 1.5406 Å) in the range 10° < 2θ < 68° at a 0.02° step size. Two configurations were chosen: Bragg-Brentano geometry with 0.15 s dwell time for the powder, and 2° glancing angle configuration with 20 s dwell time for the coating to avoid the signal contribution from the substrate.

Wear tests were carried out using a ball on flat rotary tribometer (Ducom, The Netherlands) with a 6 mm alumina ball (Dejay ltd, United Kingdom) as countersurface. The tests were carried out on the top surface of the coating, ground with SiC grinding discs (grit size P240, P400, P800 and P1200) and polished with 6 µm and 1 µm diamond pads. The resulting surface roughness $R_a$ of the samples was measured to be: $R_a < 1.1$ µm. The loads applied were 10, 20, 30 and 40 N, and the tests ran for 30 minutes at 60 RPM along a 12 mm diameter circular wear track for a total of 1800 cycles, or 68 m. Each of the test was repeated twice and the standard error from the two was associated with the specific wear rate values. The coefficient of friction graphs presented were obtained by a point-wise average of the values from each pair of repeat tests. According to Hertzian contact calculations [24], the initial maximum contact pressure ($\rho_{max}$) and the depth of maximum shear stress ($\tau_{max}$) were calculated and are reported in Table 1. It is worth noting that the depth of maximum shear stress is always below the coating/substrate interface due to the coating thickness being lower, therefore this is not expected to influence the relative wear behaviour between the different loads.

**Table 1 – Hertzian contact** - Initial maximum contact pressure and depth of maximum shear stress calculated for the different wear test loads.

|  | 10 N | 20 N | 30 N | 40 N |
| --- | --- | --- | --- | --- |
| $\rho_{max}$ (GPa) | 1.24 | 1.57 | 1.79 | 1.97 |
| $\tau_{max}$ (µm) | 31 | 39 | 45 | 49 |

Profilometry was carried out using an Alicona 5G XL (Bruker, Germany) with a 10X objective, yielding a lateral resolution of 2 µm and a vertical resolution of 50 nm. The software IF-MeasureSuite (Bruker, Germany) and the software MountainsMap (Digital Surf, France) were used for the 3D view images and for the metrology measurements, respectively. The cross-sectional profile of the wear track was measured in 4 positions along the wear track. The wear volume loss of the coating was calculated by multiplying the average cross-sectional area by the wear track length. The wear volume loss of the ball was calculated assuming the removal of a spherical cap equivalent, as explained in [25], and the specific wear rate was then calculated by dividing by the load and distance.

### 3 - Results and discussion

### 3.1 - Coating microstructure

The cross section of the coating is shown in Figure 3. A (20.9 ± 0.9) µm thick coating was obtained, as measured by averaging 10 measurements and presenting the standard error as uncertainty. The coating shows a good coating/substrate interface. The SEM BSE image in Figure 3b helps to identify the different phases present. According to their contrast level, where brighter indicates a higher atomic density, it is possible to find YAG (bright areas - circle), SiC (grey areas - square) and porosity (dark areas - rhombus).

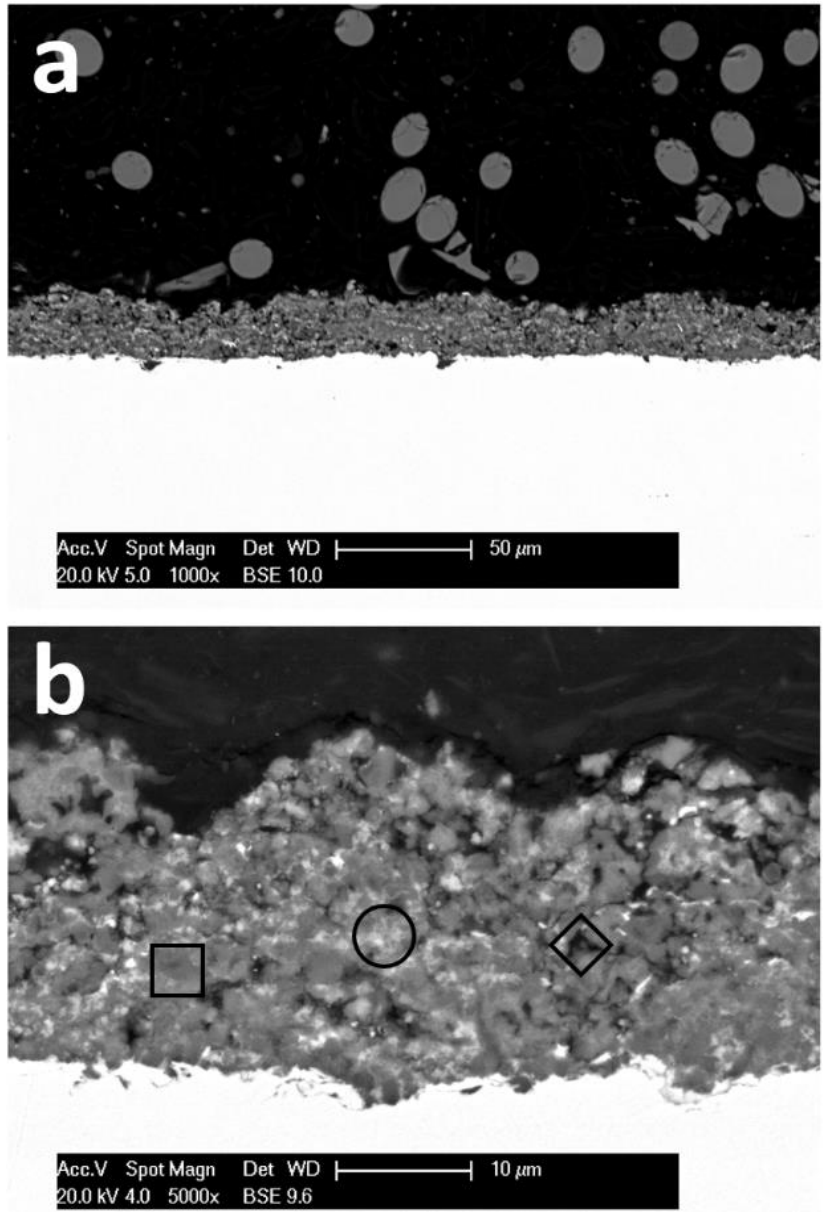

**Figure 3 – Coating cross section –** Low (a) and high (b) magnification SEM BSE images of the cross section of the coating. In (b), the circle indicates YAG rich areas, the square areas rich in SiC and the rhombus indicates porosity.

### 3.2 - Porosity and microhardness

The coating porosity was measured to be (1.6 ± 0.4) %. This porosity value is comparatively lower than that reported in the literature for similar coatings [10]. The mechanical properties of the coating were evaluated by measuring its microhardness. The coating microhardness was (300 ± 19) $HV_{0.025}$. This value is much lower than the bulk value of (2214 ± 69) $HV_{2.0}$, but comparatively only slightly lower than other similarly obtained coatings for which it was (484 ± 47 $HV_{0.1}$) [1]. This can be explained by a lower cohesion and by the presence of the YAG as binding matrix, which are known to yield lower microhardness values. The lower momentum and heat transfer between the flame and the feedstock from radial injection compared to axial injection can be the reason for lower cohesion. These factors contribute to hinder the oxidation of the SiC, but at the same time favour a decrease of the cohesion of the coating.

### 3.3 - Phase composition

The crystal structure and composition of the coating was studied by X-Ray Diffractometry, as shown in Figure 4.

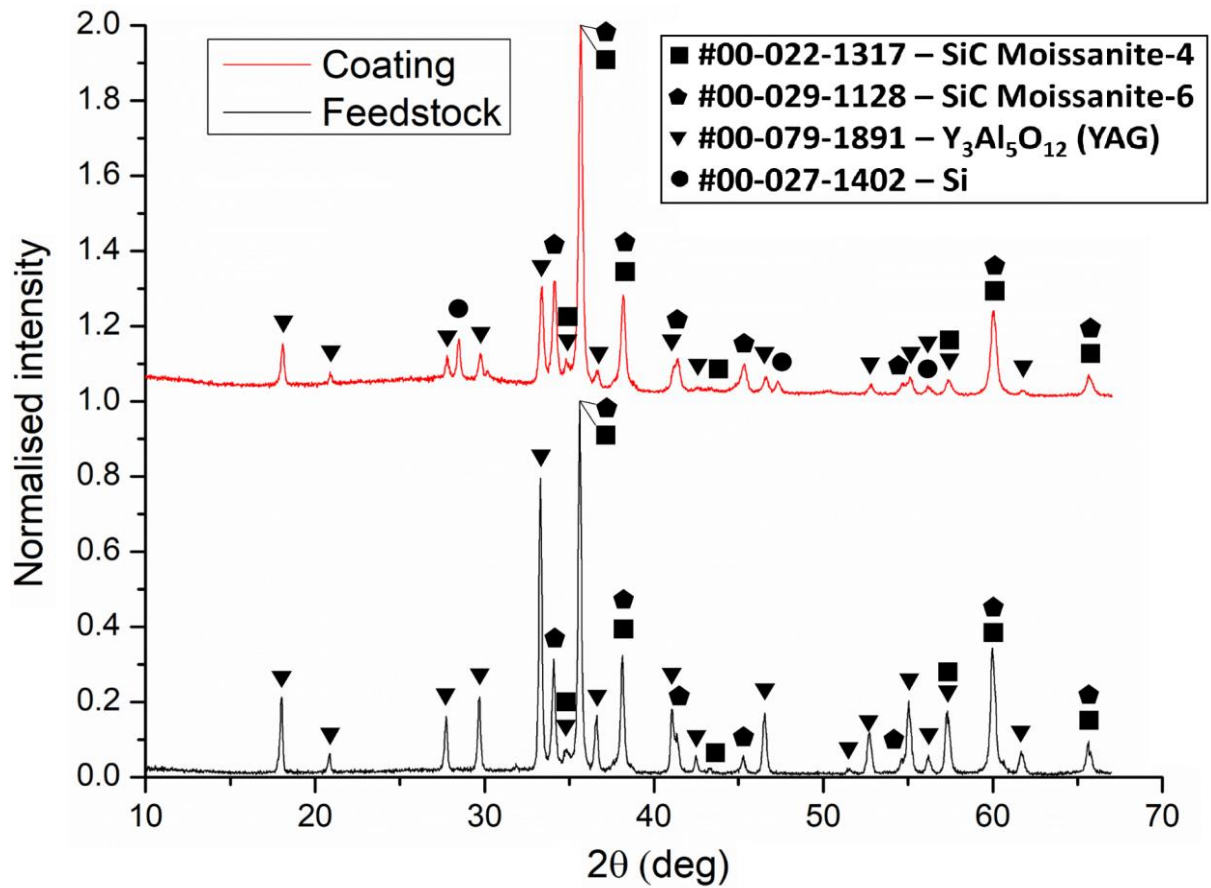

**Figure 4 – X-Ray Diffraction** – Coating (top) and feedstock (bottom) diffractograms, showing SiC and YAG in the feedstock, and additional Si in the coating, with no presence of $SiO_2$.

The diffractogram from the powder feedstock shows the presence of SiC, both in Moissanite-4 (PDF #00-073-1664) and Moissanite-6 (PDF #00-074-1302) form, and YAG (PDF #00-079-1891). Moissanite 4 and 6 are two polytypes of SiC which contain 8 and 12 atoms in the unit cell, and are therefore indicated by hP8 and hP12 Pearson symbol, respectively [26]. The diffractogram from the coating confirms the presence of these same materials; in addition, elemental Si (PDF #00-027-1402) appears as a new phase in the coating. The possibility of peaks attributed to Si being generated by other likely-present phases like $Y_2O_3$ or $YAlO_3$ has been discarded as their main peaks are not present and none of their peaks overlaps enough with the ones found experimentally. According to the Reference Intensity Ratio (RIR) method ($I/I_c$) [27], the relative concentration of the various phases was calculated semi-quantitatively and is listed in Table 2. Regarding SiC, an initial equal relative quantity of Moissanite-4 and 6 is observed in the feedstock, which then changes into having mainly Moissanite-6 in the coating, hinting polytypic transformation has occurred during the spray. In addition, a slight decrease in YAG relative content is observed, along with the appearance of a 4 wt. % of Si. The high-temperature flame causes some of the SiC in the feedstock to decompose, but the oxygen-depleted environment provided by the Hybrid Nozzle thanks to the longer expansion nozzle and the inert gas shroud offer effective oxidation prevention, leaving Si in elemental form. The partial transformation of SiC into Si through decomposition, along with the lower deposition efficiency of YAG compared to SiC, explain the relative decrease of YAG concentration in the coating compared to the feedstock. The effect of phase composition on mechanical properties of the coating is expected to be mainly determined by the dominant SiC phases; the transformation of SiC into Si is expected to have a minimal effect due to the very low final concentration if Si.

**Table 2 – Semi-quantitative XRD** - Relative concentration of the different phases as measured by XRD and calculated by the RIR method.

| Concentration (wt. %) | SiC (Moissanite-6) | SiC (Moissanite-4) | Si | YAG |
|---|---|---|---|---|
| Coating | 74 | 11 | 4 | 11 |
| Feedstock | 43 | 43 | 0 | 14 |

The presence of elemental Si was confirmed by WDS, as shown in the top-surface SEM images in Figure 5. The decomposition of SiC into the constituent elements Si and C results in areas that are stoichiometrically rich in Si.

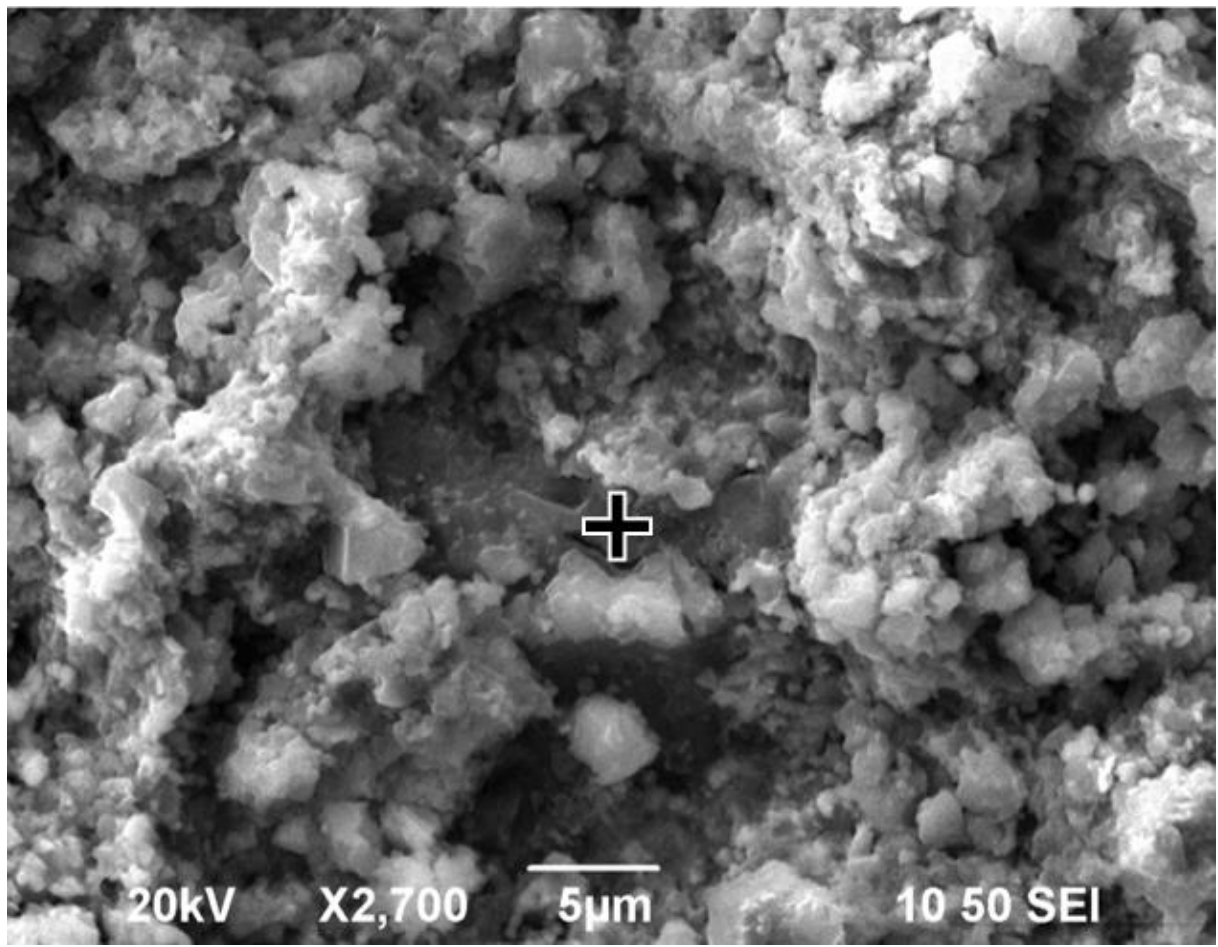

**Figure 5 – WDS –** Top surface SEM SE image from the area where the WDS spectrum was taken. The exact position is marked by a cross.

This was quantitatively assessed, and the WDS/EDS results are presented in Table 3. Here it emerges how the spectrum describes a Si rich area. The stoichiometry of this area has concentrations Si which cannot be accounted for by the formation of $SiO_2$ or the presence of residual SiC alone, but hint at the decomposition of SiC into Si and C. The residual part of Si is the result of the decomposition process, with small scattered amounts of it alone. The part of Si which is bonded with oxygen is a small fraction of the overall coating as it is below detectability in XRD. Also, it could be found only on the top surface, where it was able to react with the environment air post-deposition. The carbon residue from the dissociation process had mostly volatilised.

**Table 3 – WDS** - Relative concentration of the different phases as measured by WDS (C, O and Si elements only) and EDS (all other elements).

| Element | Signal Type | Spectrum |       |
|---------|-------------|----------|-------|
| C       | WDS         | 16.251   | at. % |
| O       | WDS         | 7.5796   | at. % |
| Al      | EDS         | 0.2779   | at. % |
| Si      | WDS         | 75.0927  | at. % |
| Cr      | EDS         | 0.1714   | at. % |
| Mn      | EDS         | 0.0129   | at. % |
| Fe      | EDS         | 0.5173   | at. % |
| Ni      | EDS         | 0.0463   | at. % |
| Cu      | EDS         | 0.051    | at. % |

A possible explanation for the formation mechanism of Si here follows. The environment in which the kerosene combustion with oxygen occurs is described by the reaction (1).

$$2\ C_{12}H_{26}\ (l) + 37\ O_2\ (g) \rightarrow 24\ CO_2\ (g) + 26\ H_2O\ (g) \quad (1)$$

The point at which the feedstock is injected is well-after the combustion chamber, therefore the above reaction is already completed at the feedstock injection point. This leaves a high-temperature, $CO_2$-rich environment with which SiC can react, and the reactions describing its decomposition are the following reactions (2) and (3).

$$SiC\ (s) + CO_2\ (g) \rightarrow Si\ (s) + 2\ CO\ (g) \quad (2)$$

$$SiC\ (s) + CO_2\ (g) \rightarrow Si\ (s) + 2\ CO\ (g) \quad (3)$$

Also, thermal decomposition of SiC at high temperature has been observed in plasma spray [28] and can be expressed by reaction (4):

$$SiC\ (s) \rightarrow Si\ (s) + C\ (s) \quad (4)$$

Thermal decomposition is less likely in our HVOLF system due to the lower temperature involved compared to Plasma spray, but temperature can still be as high as 2400 K [29]

The newly dissociated Si could then react again with $CO_2$ according to the following reactions (5) and (6)

$$SiC\ (s) + 2CO_2\ (g) \rightarrow SiO_2\ (s) + 2\ CO\ (g) \quad (5)$$

$$SiC\ (l) + 2CO_2\ (g) \rightarrow SiO_2\ (s) + 2\ CO\ (g) \quad (6)$$

Additionally, the injection of an inert gas ($N_2$) prevents oxygen entrainment from the atmosphere to the thermal spray jet. Therefore, other reactions of Si with O as (7) would be hindered in-flight and more likely at the substrate which is located in ambient environment.

$$Si\ (s) + O_2\ (g) \rightarrow SiO_2\ (s) \quad (7)$$

This possibility cannot be ruled out, however no evidence of $SiO_2$ was found by XRD; if this had occurred layer-by-layer among subsequent passes, the amount of $SiO_2$ must be either in amorphous form or crystalline but below the detectability threshold of 2 wt. % concentration. $SiO_2$ formation was however suggested by the WDX acquired from the coating's top surface, hinting that at least after the last pass some oxide formation is possible on the surface.

**3.4 - Wear performance**

The wear performance of the coatings was tested against an alumina counterbody in order to provide a ceramic-on-ceramic wear couple. The results are shown in Figure 6. The volumetric specific wear rate is shown in Figure 6a. The wear loss of the coating generally appears higher than that of the counterbody. Also, for the coating, an overall increase of the wear loss occurs as the load increases, whereas for the counterbody, the values remain very similar at all loads. The wear loss increase of the coating appears to be steeper between 20 N and 30 N loads, and also the variability as shown by the error bars increases at the same loads, suggesting the inception of cracking wear mechanism of the ceramics, as will be analysed in Section 3.5. The coefficient of friction at the various loads as a function of wear distance is shown in Figure 6b. The running-in period appears very different between the 10 N test and all the other tests, with the former starting at a low value of 0.2 and gradually increasing up to a stable value along with the whole wear distance, and the latter increasing steeply up to values above 0.5 and stabilising as early as after 20 m. The highest steady-state coefficient of friction value is shown by the 20 N load test, with lower values shown by 20 N and 40 N loads. Overall, the coating shows a remarkably lower coefficient of friction at low load (≤ 10 N), suggesting its best suitability for low-load applications. The coefficient of friction values are similar to those reported for similar SiC/30%YAG coatings against ceramic (SiC) counterbody in unlubricated conditions [2], but it is noteworthy that the test conditions are different in terms of counterbody and testing rig. Regarding wear loss, the specific wear rate is comparatively higher than that measured for a SiC/30%YAG coating against stainless steel counterbody, with a value of $2.5 \cdot 10^{-5}$ mm$^3$/Nm here compared to $1.9 \cdot 10^{-6}$ mm$^3$/Nm in [1], notwithstanding their higher initial contact pressure of 1.6 GPa. To complete the picture, the results in this work can be compared with those obtained for the same counterbody, alumina, against pure SiC obtained by sintering, however, at a higher load (214 N) [30]. In that case, the coefficient of friction and the wear loss values appeared larger than those obtained in this work. The applied load and counterbody material have certainly a prominent role in determining the wear performance. Nonetheless, it is noteworthy how the amount of binder material (YAG), which in our case is at a value in between (11 wt. %), may account for the explanation of the better wear performance as it can reduce SiC cracking and grain pull-out compared to the pure SiC used in [30].

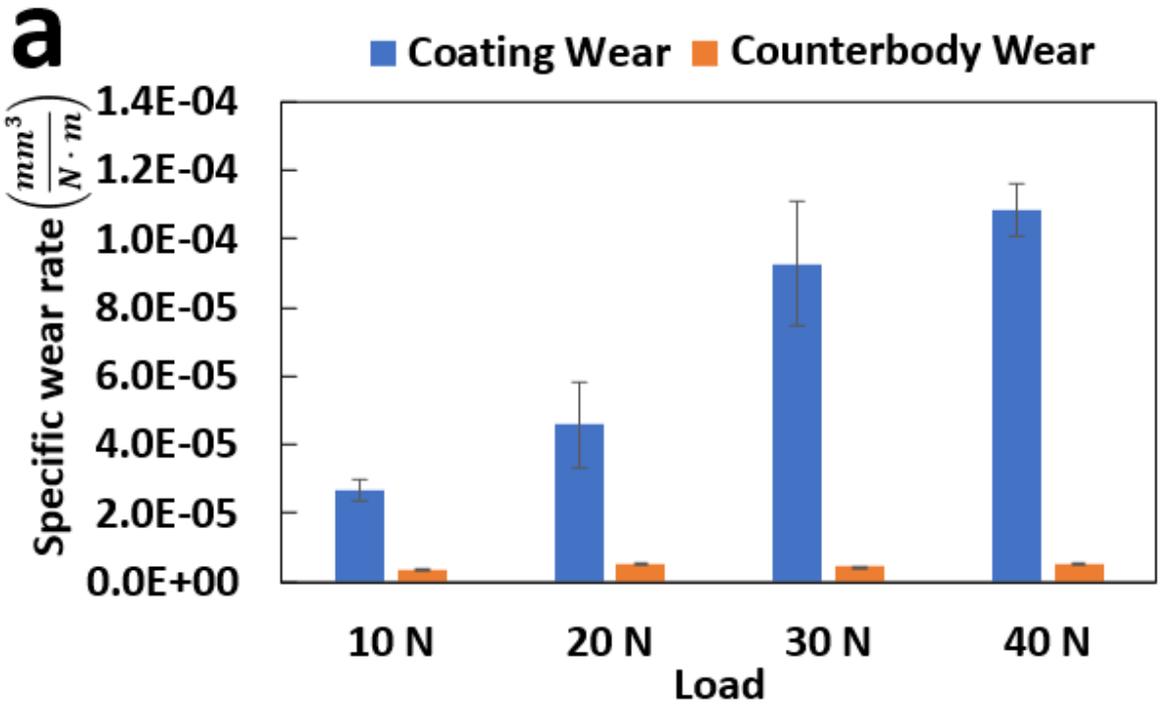

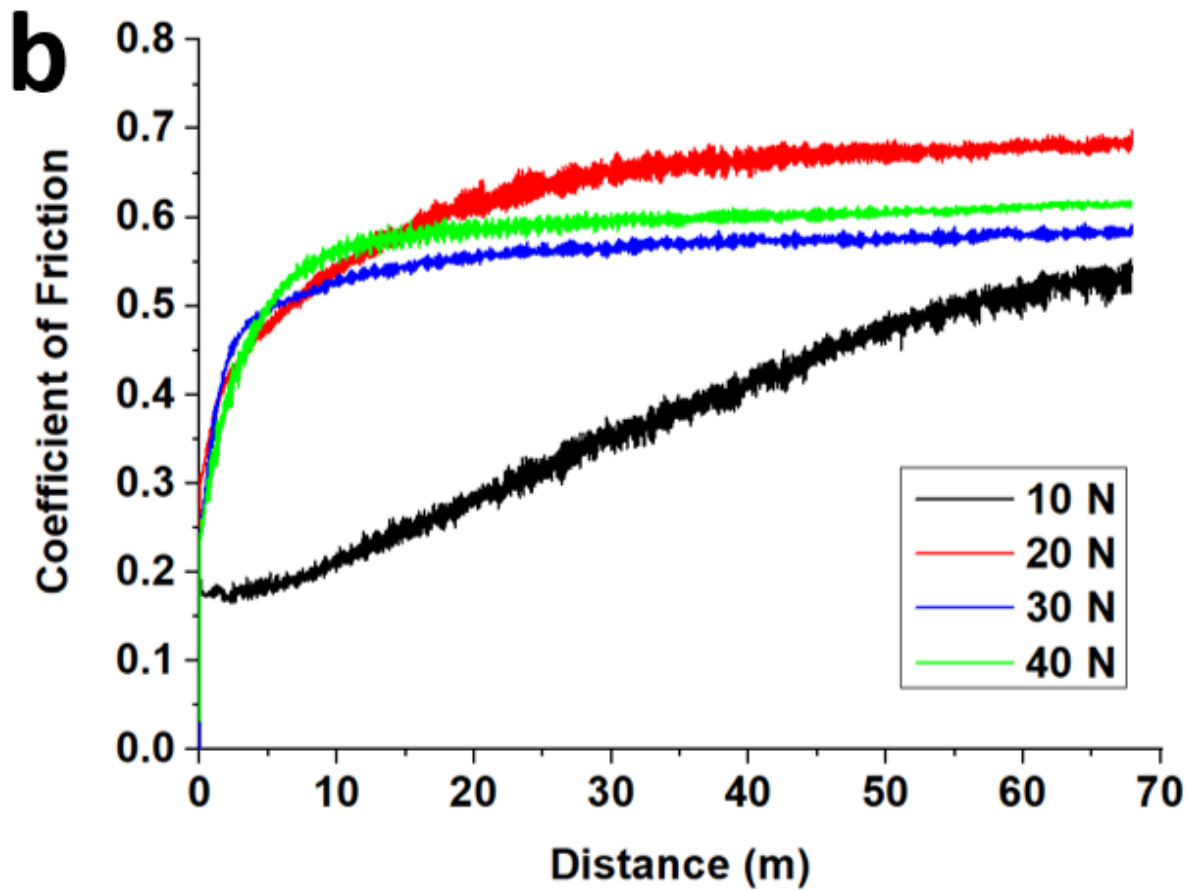

**Figure 6 – Wear tests** – Volumetric specific wear rate (a) and coefficient of friction over distance (b) of SiC coatings against alumina counterbody at 10, 20, 30 and 40 N loads.

**3.5 - Wear mechanisms**

Further insight into the wear mechanisms at the various loads is given by the SEM images in Figure 7, which shows the full-width wear tracks along with high-resolution details from within the wear tracks. The main features emerging from the low magnification SEM images are the increase of wear track width as the load increases, and especially between the 10 N load test and all the others, as reported in Table 4. Cracking is evident at all loads but is more pronounced at 40 N. Some localised cracking is apparent also outside the wear track, hinting it was present before the test and may have resulted from the grinding and polishing procedure. The high magnification images add more details to this picture. At 10 N, a mixture of flat, deformed areas coexists with rougher areas full of debris. Even if the running-in period, which was characterised by a lower coefficient of friction, had already taken place, some parts of the coating top surface were still to be flattened by the counterbody, and the debris tended to be stored in those areas. At the higher load of 20 N, the wear track appears more uniformly flattened, with a smooth, slightly cracked surface, and little debris. This is the highest load before large-scale cracking mechanisms occur to release the stress of the coating. The higher loads of 30 and 40 N cause relatively more cracking and debris that helps keep the coefficient of friction lower due to a three-body friction. The 10 N load test shows a different behaviour because it is so mild that the running-in deformation of the coating top surface takes the whole test length to occur. The higher coefficient of friction observed at 20 N can be explained by the larger contact surface given by the smoother finish and lack of debris. At 30 N, the cracking increases slightly, and much more debris is forming. The 40 N test is the harshest scenario, with large-scale and micro cracking, debris, ploughing and grain pull-out. The micro cracking appears as small wrinkles, the ploughing as long horizontal trails across the image, and the grain pull-out appears along the crack at the bottom left of the image, where the right crack border appears brighter due to the weaker electrical contact from the pull-out.

**Table 4 – Wear track width** – Width of the wear track at the various loads tested. The width was measured at 10 points along the wear track and averaged, and is presented along with the standard error.

| Wear test load | 10 N | 20 N | 30 N | 40 N |
|---|---|---|---|---|
| Wear track width | (308 ± 11) μm | (806 ± 10) μm | (865 ± 8) μm | (1004 ± 8) μm |

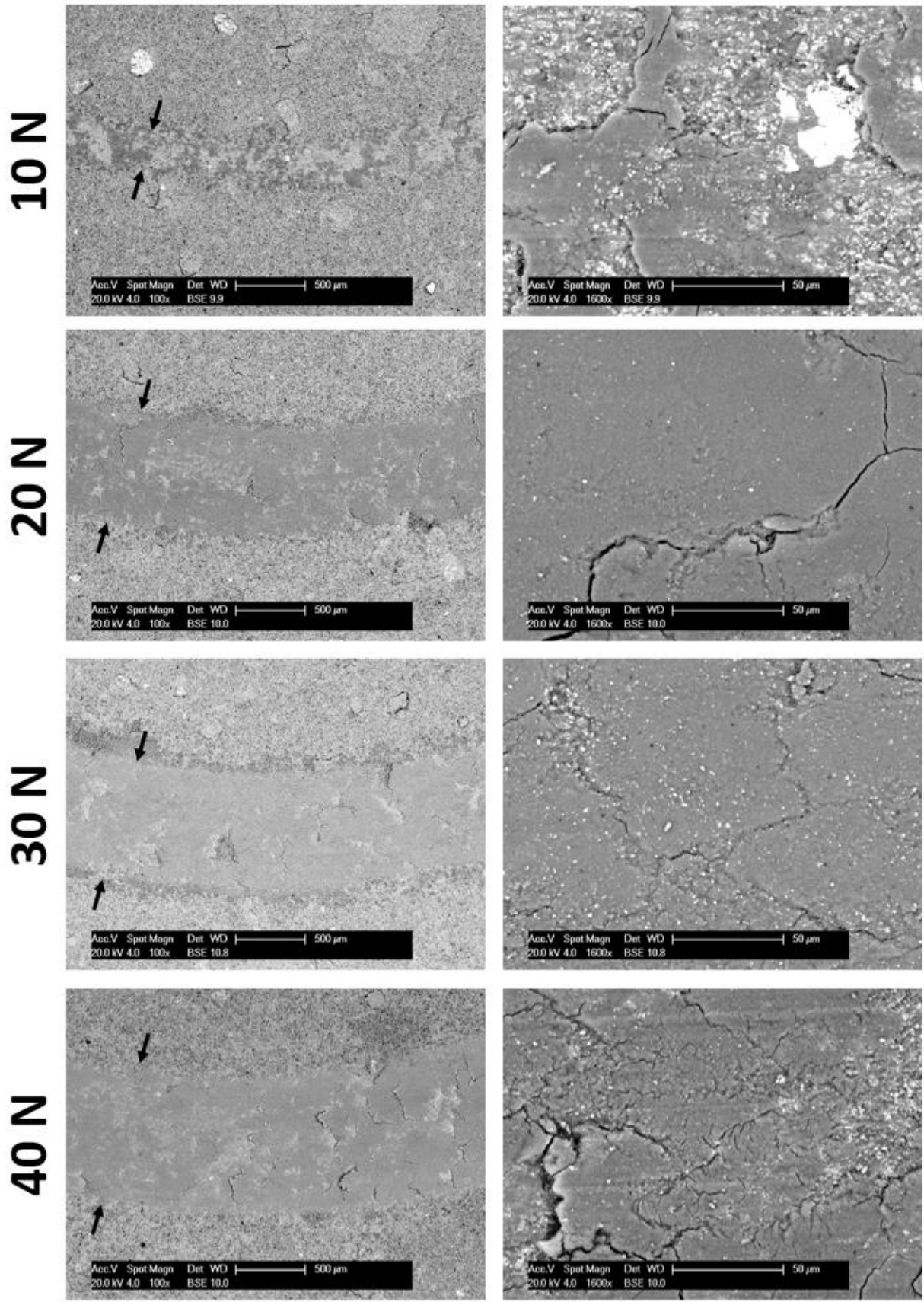

**Figure 7 – Worn surface SEM** – SEM BSE images of the wear tracks at low magnification (left) and high magnification (right). The arrows on the left-hand images highlight the location and width of the wear track.

A 3D view of the wear tracks at the various loads is shown in Figure 8. Here, similar features to those previously shown are present, as the track width and cracking increase with the load. In addition, it is possible to observe the depth profile of the wear track deepening as the load increases, by observing the image boundary bending at the top of each image.

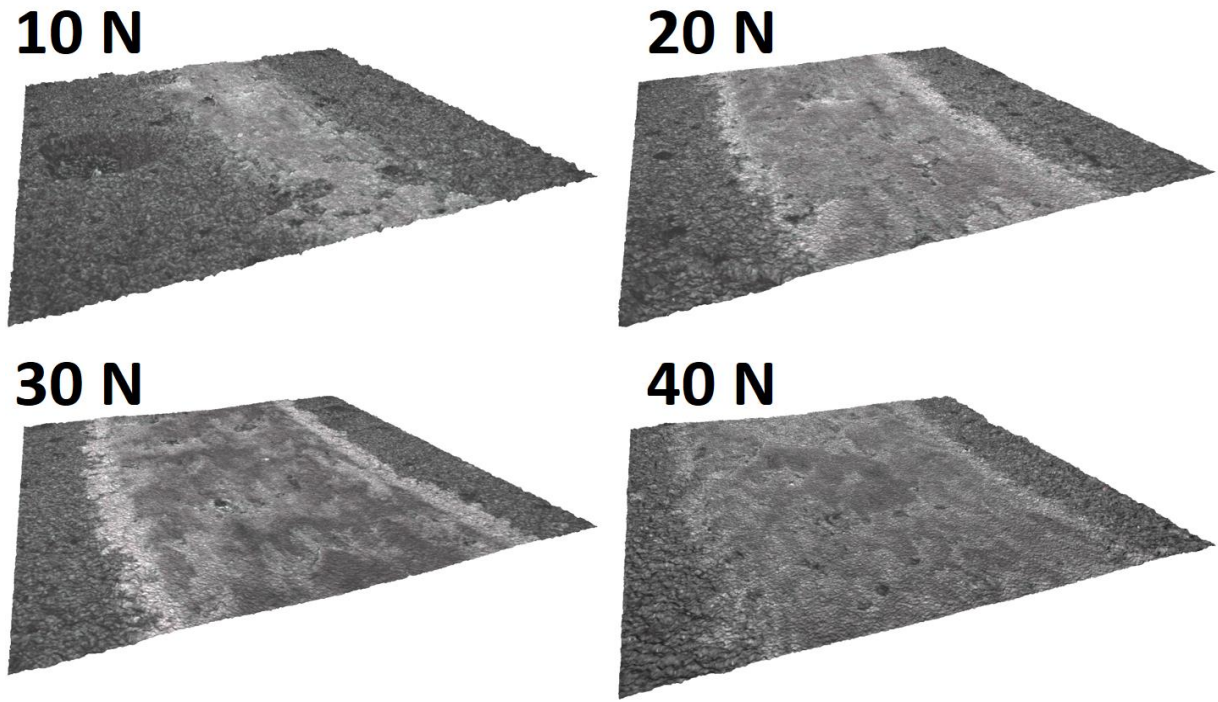

**Figure 8 – Worn surface profilometry –** 3D views of the wear track at the various loads, showing how the depth profile deepens as the load increases, allowing to appreciate the scale of the wear damage at the different loads. The depth profile is exaggerated by a factor 2 to make the depth profile better visible. The areas presented are squares of side 2 mm.

Adding to this picture, Figure 9 shows the worn surface of the counterbody ball when tested at various loads. It can be seen how ploughing emerges as the main wear mechanism indicating strong abrasive wear taking place – as indicated by the threads carved along the wear direction. This is aided by the presence of wear debris, which causes three-body abrasive wear to occur. No major material transfer between the ball and the coating is visible, indicating debris is mostly displaced out of the wear track. The dominant ceramic character of the coating and counterbody lead to a brittle and non-ductile behaviour, for which the wear debris tends not to be incorporated within the mating surfaces.

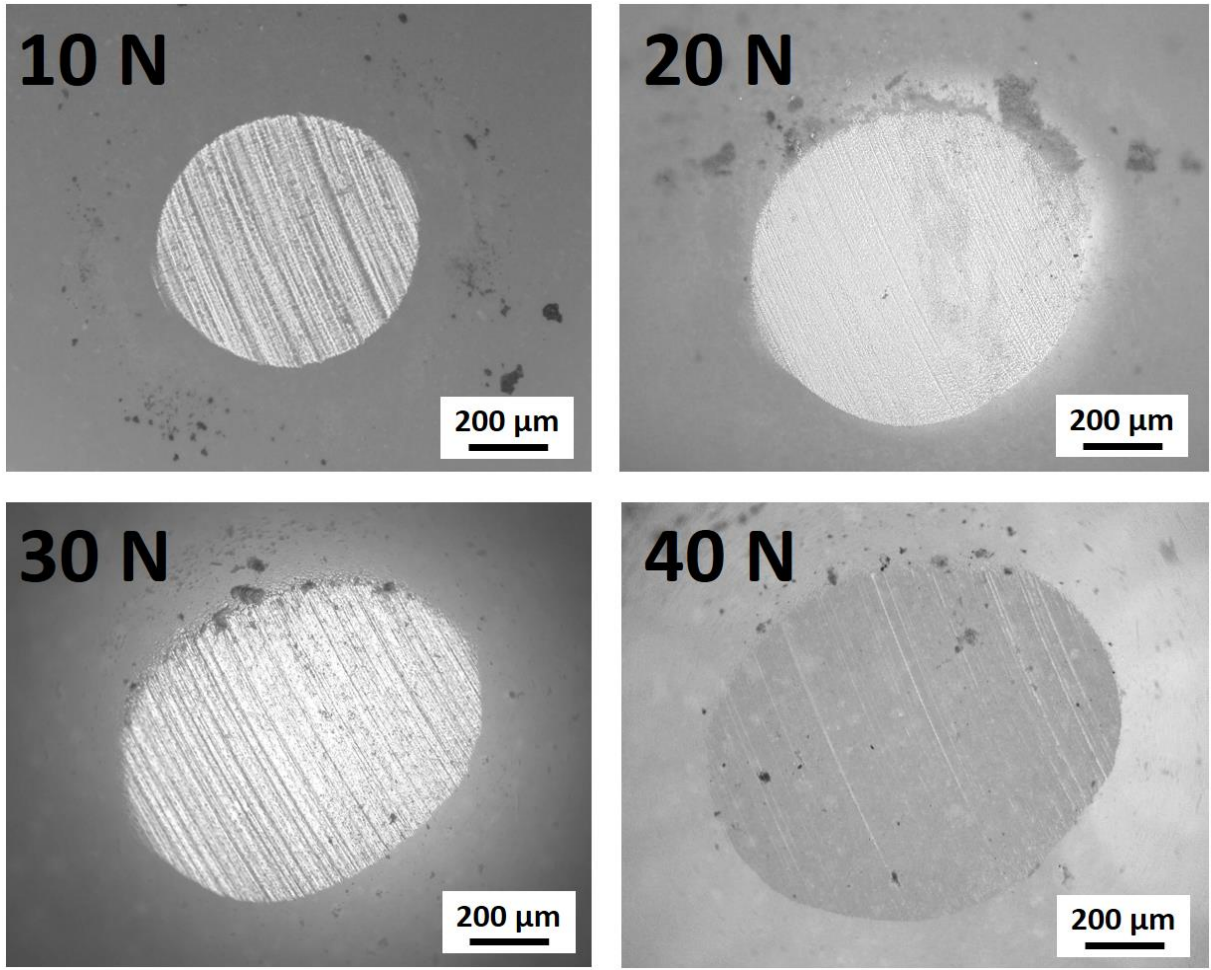

**Figure 9 – Counterbody wear –** Optical microscopy images of the $Al_2O_3$ counterbody ball worn surface at the various loads.

The wear mechanism at high load (40 N) is better reported in Figure 10. Here, a detailed 3D view of a cracking event associated with grain pull-out is shown. Traversing the image from the top left to the bottom right, along the counterbody direction of travel, one encounters a depression (1), a bulge (2), a vertical crack (3) and the grain pull-out and removal of the grain from the coating surface (4). This wear mechanism occurs at the highest load tested (40 N) and will be the main responsible for the gradual failure of the coating. Milder wear mechanisms, such as ploughing, are also visible in the darker trails traversing the image along the counterbody direction of travel.

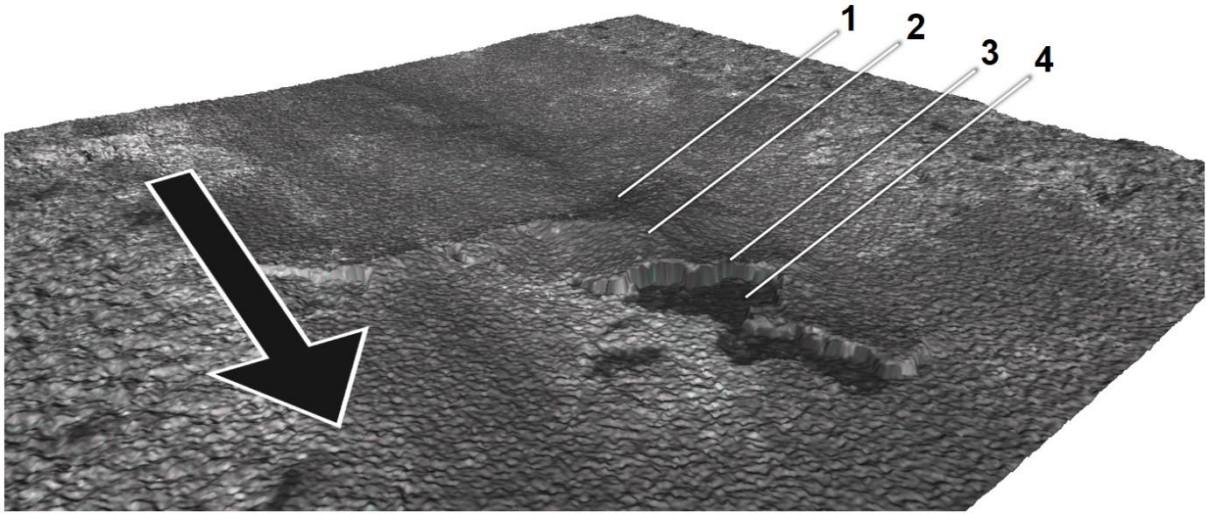

**Figure 10 – Wear mechanism at 40 N –** Detail of the 40 N wear track showing a 3D view of a cracking and pull-out event. The arrow indicates the direction of the counterbody during the wear test. A depression (1) and a bulge (2) are present as part of the same wear mechanism. The cracking (3) and pull-out (4) can be seen following the direction of travel of the counterbody. The height profile is exaggerated by a factor 2 to make the depth profile better visible. The area presented is extracted from a square of side 2 mm.

### 3.6 – Effect of the Hybrid Nozzle setup on the coating

The effect of several factors has contributed to the microstructure, properties and performance of this SiC/YAG coating. The choice of a liquid-fuelled HVOF thermal spray torch, a standard in the industry for wear resistant coating, yields a particularly powerful flame in HVOF thermal spray (~250 kW) characterised by high-velocity and moderate temperature. The solid shroud placed at the exit of the torch allows a constrained expansion towards the environment, modifying the thermodynamics of the flame. The use of radial injection with this kind of system allows exploiting a fraction of this power by letting the feedstock interact with the thermal spray flame within this expansion section, easily injecting the feedstock inside the flame and allowing the mixing and the heat and momentum transfer. The choice of liquid feedstock – suspension in this case – makes it easier to handle fine powder that would not be suitable otherwise. The water in the suspension requires to be accelerated and vaporised by the thermal spray flame, reducing the flame power by approximately 1.9 kW at 50 ml/min assuming complete acceleration and subsequent vaporisation [22]. The milder feedstock-flame interaction, along with the shrouding gas protection at the nozzle end, makes the feedstock reach lower temperatures, limiting melting and oxidation. This is, however, accompanied by a lower momentum and accordingly a lower deposition efficiency can be envisaged – but also a lower loss of SiC through decomposition. Proof of this is the lack of $SiO_2$ and the presence of elemental Si, meaning that decomposition of SiC at high temperature in the oxygen-depleted environment occurs thanks to the Hybrid Nozzle, the reducing gas ratio of the HVOLF and the YAG shell, prevent the oxygen from bonding with Si and lets it deposit in elemental form.

This work opens the way for further investigations of HVOLF with liquid feedstock, including both suspension and solution precursor feedstock. The applicability to additional heat- and oxidation-sensitive materials can be foreseen, including carbides, nitrides and nanomaterials. The flexibility of the system allows for the deposition of composites with mixing *in situ* by injecting ordinary powder feedstock within the HVOLF torch and, simultaneously, liquid feedstock through the radial injection in the Hybrid Nozzle.

## 4 – Conclusion

A novel suspension HVOLF setup comprising a solid shroud, radial feedstock injection and inert gas shrouding was used to deposit SiC/YAG coatings, allowing to draw the following conclusions:

- A SiC/YAG coating was obtained with only 11 wt.% YAG binder content, and no or minimal (< 2 wt. %) Si oxidation detectable through XRD.

- The Hybrid Nozzle attachment used in the setup allowed an oxygen-depleted environment as proved by the presence of 4 wt. % elemental Si in the coating, which had undergone decomposition but not oxidation.

- The coating wear performance showed the lowest coefficient of friction at low load (10 N) and the onset of wear mechanisms characteristic of ceramic wear in the brittle regime between 20 and 30 N, indicating the coating is performing better for low load applications, under 20 N.

- Main wear mechanisms included abrasive wear (ploughing, three-body) and grain pull-out. At the lowest load (10 N) only ploughing occurs. The lower microhardness played a role in wear performance and is most likely the consequence of lower cohesion.

**Acknowledgements**

This work was supported by the Engineering and Physical Sciences Research Council [grant number EP/V010093/1]. The authors acknowledge the Nanoscale and Microscale Research Centre (nmRC) at the University of Nottingham for access to the SEM and WDX facility, and Dr Lorelei Robertson for the contribution with WDX. John Kirk is gratefully acknowledged for the careful management of the thermal spray facilities.

**References**

[1] Mubarok F, Espallargas N. Tribological behaviour of thermally sprayed silicon carbide coatings. Tribology International. 2015;85;56-65. http://dx.doi.org/10.1016/j.triboint.2014.11.027

[2] Mubarok F, Armada S, Fagoaga I, Espallargas N. Thermally Sprayed SiC Coatings for Offshore Wind Turbine Bearing Applications. Journal of Thermal Spray Technology. 2013;22;1303-9. https://doi.org/10.1007/s11666-013-9991-y

[3] Acheson EG. US Patent no US568323A. 1896

[4] Krstic VD. Production of Fine, High-Purity Beta Silicon Carbide Powders. Journal of the American Ceramic Society. 1992;75;170-4. https://doi.org/10.1111/j.1151-2916.1992.tb05460.x

[5] Lely JA. Darstellung von Einkristallen von Silicium Carbid und Beherrschung von Art und Menge der eingebauten Verunreinigungen. Berichte der Deutschen Keramischen Gesellschaft. 1955;32;229-36.

[6] Hayun S, Paris V, Mitrani R, Kalabukhov S, Dariel MP, Zaretsky E, et al. Microstructure and mechanical properties of silicon carbide processed by Spark Plasma Sintering (SPS). Ceramics International. 2012;38;6335-40. https://doi.org/10.1016/j.ceramint.2012.05.003

[7] Lilov SK, Study of the equilibrium processes in the gas phase during silicon carbide sublimation. Materials Science and Engineering: B. 1993;21;65-9. https://doi.org/10.1016/0921-5107(93)90267-Q

[8] Ervin Jr G. Oxidation Behavior of Silicon Carbide. Journal of the American Ceramic Society. 1958;41;347-52. https://doi.org/10.1111/j.1151-2916.1958.tb12932.x

[9] Balat MJH. Determination of the active-to-passive transition in the oxidation of silicon carbide in standard and microwave-excited air. Journal of the European Ceramic Society. 1996;16;55-62. https://doi.org/10.1016/0955-2219(95)00104-2

[10] Mubarok F, Espallargas N, Suspension Plasma Spraying of Sub-micron Silicon Carbide Composite Coatings. Journal of Thermal Spray Technology. 2015;24;817-25. https://doi.org/10.1007/s11666-015-0242-2

[11] Tului M, Giambi B, Lionetti S, Pulci G, Sarasini F, Valente T. Silicon carbide based plasma sprayed coatings. Surface and Coatings Technology. 2012;207;182-9. https://doi.org/10.1016/j.surfcoat.2012.06.062

[12] Hu C, Niu Y, Li H, Ren M, Zheng X, Sun J. SiC Coatings for Carbon/Carbon Composites Fabricated by Vacuum Plasma Spraying Technology. Journal of Thermal Spray Technology. 2012;21;16-22. https://doi.org/10.1007/s11666-011-9697-y

[13] Miranda FS, Caliari FR, Campos TM, Essiptchouk AM, Filho GP. Deposition of graded $SiO_2$/SiC coatings using high-velocity solution plasma spray. Ceramics International. 2017;43;16416-23. https://doi.org/10.1016/j.ceramint.2017.09.018

[14] Lee HY, Yu YH, Lee YC, Hong YP, Ko KH. Cold spray of SiC and $Al_2O_3$ with soft metal incorporation: A technical contribution. Journal of Thermal Spray Technology. 2004;13;184-9. https://doi.org/10.1361/10599630419355

[15] Fals HC, Aguiar D, Fanton L, Belém MJX, Lima CRC. A new approach of abrasive wear performance of flame sprayed NiCrSiBFeC/SiC composite coating. Wear. 2021;477;203887. https://doi.org/10.1016/j.wear.2021.203887

[16] Rincón A, Pala Z, Hussain T. A suspension high velocity oxy-fuel thermal spray manufacturing route for silicon carbide – YAG composite coatings. Materials Letters. 2020;281;128601. https://doi.org/10.1016/j.matlet.2020.128601

[17] Memon H, Rincón Romero A, Derelizade K, Venturi F, Hussain T. A new hybrid suspension and solution precursor thermal spray for wear resistant silicon carbide composite coatings. Materials & Design. 2022;224;111382. https://doi.org/10.1016/j.matdes.2022.111382

[18] Thomson I, Pershin V, Mostaghimi J, Chandra S. Experimental Testing of a Curvilinear Gas Shroud Nozzle for Improved Plasma Spraying. Plasma Chemistry and Plasma Processing. 2001;21;65-82. https://doi.org/10.1023/A:1007041428743

[19] Dolatabadi A, Mostaghimi J, Pershin V. Effect of a cylindrical shroud on particle conditions in high velocity oxy-fuel (HVOF) spray process. Journal of Materials Processing Technology. 2003;137;214-24. https://doi.org/10.1016/S0924-0136(02)01084-1

[20] Toma F-L, Sokolov D, Bertrand G, Klein D, Coddet C, Meunier C. Comparison of the photocatalytic behavior of $TiO_2$ coatings elaborated by different thermal spraying processes. Journal of Thermal Spray Technology. 2006;15;576-81. https://doi.org/10.1361/105996306X147225

[21] Venturi F, Rance GA, Thomas J, Hussain T. A low-friction graphene nanoplatelets film from suspension high velocity oxy-fuel thermal spray. AIP Advances. 2019;9;025216. https://doi.org/10.1063/1.50890211

[22] Venturi F, Hussain T. Radial injection in suspension high velocity oxy-fuel (S-HVOF) thermal spray of graphene nanoplatelets for tribology. Journal of Thermal Spray Technology. 2020;29;255-69. https://doi.org/10.1007/s11666-019-00957-y

[23] Chadha S, Jefferson-Loveday R, Venturi F, Hussain T. A Computational and Experimental Investigation into Radial Injection for Suspension High Velocity Oxy-Fuel (SHVOF) Thermal Spray. Journal of Thermal Spray Technology. 2019;28;1126-45. https://doi.org/10.1007/s11666-019-00888-8

[24] Budynas R, Nisbett K. Shigley's Mechanical Engineering Design. 10th edition. 2014. McGraw-Hill, New York.

[25] Pulsford J. An investigation of cermet and composite HVOF thermal spray coatings for internal surfaces. 2020. EngD thesis, University of Nottingham. http://eprints.nottingham.ac.uk/id/eprint/63540

[26] Hannam AL, Shaffer PTB. Revised X-ray diffraction line intensities for silicon carbide polytypes, Journal of Applied Crystallography. 1969;2;45-48. https://doi.org/10.1107/S0021889869006510

[27] Hubbard CR, Snyder RL. RIR - Measurement and Use in Quantitative XRD. Powder Diffraction. 1988;3;74-7. https://doi.org/10.1017/S0885715600013257

[28] Kang HK, Kang SB. Thermal decomposition of silicon carbide in a plasma-sprayed Cu/SiC composite deposit. Materials Science and Engineering: A. 2006;428;336-45. https://doi.org/10.1016/j.msea.2006.05.054

[29] Kamnis S, Gu S. 3-D modelling of kerosene-fuelled HVOF thermal spray gun. Chemical Engineering Science. 2006;61;5427-39. https://doi.org/10.1016/j.ces.2006.04.005

[30] Yang Q, Senda T, Kotani N, Hirose A. Sliding wear behavior and tribofilm formation of ceramics at high temperatures. Surface and Coatings Technology. 2004;184;270-7. https://doi.org/10.1016/j.surfcoat.2003.10.157